\documentclass[aps,prl,showpacs,twocolumn,superscriptaddress]{revtex4}
\usepackage{bm,color}
\usepackage{graphicx}
\usepackage{amsmath}

\begin{document}
%\draft
\title {Theory of magnetic switching of ferroelectricity in spiral magnets}
\author{Masahito Mochizuki}
%%\email{mochizuki@erato-mf.t.u-tokyo.ac.jp}
\affiliation{Department of Applied Physics, University of Tokyo,
Tokyo 113-8656, Japan}

\author{Nobuo Furukawa}
%%\email{furukawa@phys.aoyama.ac.jp}
\affiliation{Department of Physics, Aoyama Gakuin University,
Sagamihara, 229-8558 Japan}
\affiliation{Multiferroics Project, ERATO, Japan Science and Technology 
Agency (JST), Tokyo 113-8656, Japan}

\begin{abstract}
%%Ferroelectric properties under magnetic fields ($\bm H_{\rm ex}$) in 
%%several multiferroics cannot be understood simply, and their microscopic
%%understanding is an urgent issue. Here w
We propose a microscopic theory for magnetic switching of electric 
polarization ($\bm P$) in the spin-spiral multiferroics by taking 
TbMnO$_3$ and DyMnO$_3$ as examples. We reproduce 
their phase diagrams under a magnetic field $\bm H_{\rm ex}$ by 
Monte-Carlo simulation of an accurate spin model and reveal that 
competition among the Dzyaloshinskii-Moriya interaction, spin anisotropy, 
and spin exchange is controlled by the applied $\bm H_{\rm ex}$, 
resulting in magnetic transitions accompanied by reorientation or 
vanishing of $\bm P$. We also discuss the relevance of the proposed 
mechanisms to many other multiferroics such as LiCu$_2$O$_2$, MnWO$_4$, 
and Ni$_3$V$_2$O$_4$.
\end{abstract}
%% 75.80.+q Magnetomechanical and magnetoelectric effects, magnetostriction
%% 77.80.-e Ferroelectricity and antiferroelectricity
%% 77.80.Bh Phase transitions and Curie point
%% 77.80.Dj Domain structure; hysteresis
%% 77.80.Fm Switching phenomena  
%% 75.30.-m Intrinsic properties of magnetically ordered materials
%% 75.30.Gw Magnetic anisotropy  
%% 75.47.-m Magnetotransport phenomena; materials for magnetotransport
%% 75.47.Lx Manganites 
\maketitle
%\sloppy \maketitle
Concurrently magnetic and ferroelectric materials, i.e. multiferroics, 
offer prospective systems to attain magnetic control of electricity 
via magnetoelectric (ME) coupling~\cite{Kimura03a,Reviews}.
It was experimentally demonstrated that an external magnetic field 
($\bm H_{\rm ex}$) can cause reorientation, emergence, and vanishing of 
ferroelectric polarization $\bm P$ in many spin-spiral multiferroics 
such as $R$MnO$_3$ 
($R$=Tb, Dy, Eu$_{1-x}$Y$_x$, etc)~\cite{Kimura07,Kimura05,Murakawa08}, 
LiCu$_2$O$_2$~\cite{SPark07}, MnWO$_4$~\cite{Taniguchi08}, and
Ni$_3$V$_2$O$_4$~\cite{Kenzelmann06}.
These ME phenomena are currently attracting enormous interest,
and a thorough understanding of their mechanisms is an urgent issue.
However, the number of theoretical studies is very few despite many
experimental reports.
Naively, the applied $\bm H_{\rm ex}$ can determine the direction of $\bm P$ 
by controlling the conical spin structure via Zeeman coupling, but
there are many examples that do not obey this simple picture. 

In the spin-spiral multiferroics, inherent spin frustration as an origin 
of the spiral magnetism inevitably reduces the spin-exchange energy, and 
hence increases the relative importance of other tiny interactions, e.g. the 
single-ion spin anisotropy and the Dzyaloshinskii-Moriya (DM) interaction. 
Consequently, the magnetic switching of $\bm P$ in this new class of 
multiferroics is governed by their fine energy balance tuned by 
$\bm H_{\rm ex}$, which cannot be understood from a simple 
interplay between Zeeman coupling and the spin exchanges.
%%The {\it microscopic} understanding of competing interactions and 
%%resulting switching phenomena in multiferroics is essentially 
%%important for clarification of the general physics of cross-correlation 
%%phenomena and for their future applications.

In this Letter, by taking the Mn perovskites TbMnO$_3$ and DyMnO$_3$ as 
examples, we propose a microscopic theory for the magnetic 
control of $\bm P$ in the spin-spiral multiferroics.
Their puzzling $T$-$H_{\rm ex}$ phase diagrams are reproduced 
by the Monte-Carlo (MC) analysis of an accurate spin model.
Our microscopic theory reveals that the applied $\bm H_{\rm ex}$ 
controls conflicts among the spin exchanges, spin anisotropy, and 
DM interaction, resulting in magnetic transitions 
accompanied by reorientation or vanishing of $\bm P$. 
The mechanisms proposed here are relevant 
to many other spin-spiral multiferroics
such as LiCu$_2$O$_2$~\cite{SPark07}, MnWO$_4$~\cite{Taniguchi08}, 
and Ni$_3$V$_2$O$_4$~\cite{Kenzelmann06}.
We also discuss the influence of effective magnetic fields 
from rare-earth $f$ moments.
%%$\bm H_{fd}$ acting on the Mn spins via the $f$-$d$ coupling.

\begin{figure}[tdp]
\includegraphics[scale=1.0]{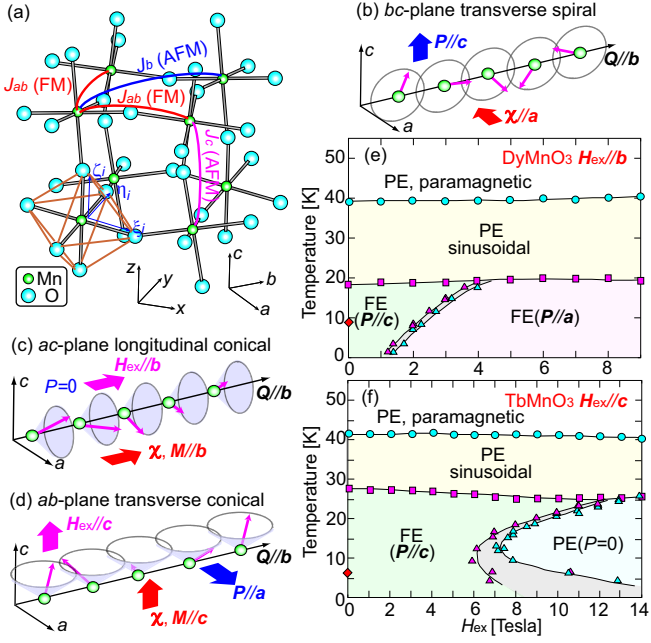}
\caption{(color online). (a) Crystal structure, spin exchanges,
%%Dzyaloshinskii-Moriya vectors $\bm d_{i,j}$
and local axes $\xi_i$, $\eta_i$, and $\zeta_i$ in $R$MnO$_3$. 
Here FM (AFM) denotes (anti)ferromagnetic exchange.
(b) $bc$-plane transverse spin spiral in 
TbMnO$_3$ and DyMnO$_3$, which induces ferroelectric polarization 
$\bm P$$\parallel$$\bm c$.
(c) [(d)] Application of $\bm H_{\rm ex}$$\parallel$$\bm b$ 
[$\bm H_{\rm ex}$$\parallel$$\bm c$] is expected
to stabilize the longitudinal [transverse] spin spiral with magnetization 
$\bm M$$\parallel$$\bm H_{\rm ex}$ where $\bm P$=0 
[$\bm P$$\parallel$$\bm a$] is expected within the spin-current model. 
(e)[(f)] Experimental $T$-$H_{\rm ex}$ phase 
diagram of DyMnO$_3$ [TbMnO$_3$] for $\bm H_{\rm ex}$$\parallel$$\bm b$ 
[$\bm H_{\rm ex}$$\parallel$$\bm c$]
from Ref.~\cite{Kimura05}, which shows reorientation of $\bm P$ from 
$\bm P$$\parallel$$\bm c$ to $\bm P$$\parallel$$\bm a$ 
[disappearance of $\bm P$]~\cite{Kimura05}. 
Here FE (PE) denotes ferroelectric (paraelectric) phase.}
\label{Fig01}
\end{figure}

The ferroelectricity in these materials is described by the 
spin-current model~\cite{Mostovoy06,Katsura05}
 as given by $\bm P \propto \bm Q \times \bm \chi$,
%%%%%%%%%%%%%%%%%%%%%%%%%%%%%%%%
%%\begin{equation}
%%\bm P \propto \bm Q \times \bm \chi,
%%\end{equation}
%%%%%%%%%%%%%%%%%%%%%%%%%%%%%%%%
where $\bm Q$ is a propagation vector of the spiral and 
$\bm \chi \propto \sum_{<i,j>} \bm S_i \times \bm S_j$ is the vector 
spin chirality.
As shown in Fig~\ref{Fig01}(b), the Mn spins in TbMnO$_3$ and DyMnO$_3$
are rotating within the $bc$ plane ($\bm \chi$$\parallel$$\bm a$) to form 
a transverse spiral with $\bm Q$$\parallel$$\bm b$~\cite{Kenzelmann05}, 
and thus $\bm P$$\parallel$$\bm c$ is realized.

%%%%%%%%%%%%%%%%%%%%%%%%%%%%%%%%%%%%%%%%%%%%%%%%%%%%%%%%%%%%%
%%%%%%%%%%%%%%%%%%%%%%%%%%%%%%%%%%%%%%%%%%%%%%%%%%%%%%%%%%%%%%
In Figs.~\ref{Fig01}(c)-(f), we briefly summarize the puzzles in 
$R$MnO$_3$~\cite{Kimura05}.
The applied $\bm H_{\rm ex}$ induces the magnetization 
$\bm M$$\parallel$$\bm H_{\rm ex}$ via Zeeman coupling, 
and hence forces the spin structure
to be conical where $\bm \chi$$\parallel$$\bm H_{\rm ex}$.
When we apply $\bm H_{\rm ex}$$\parallel$$\bm Q$
[see Fig.~\ref{Fig01}(c)], 
we expect a longitudinal conical spin order with 
$\bm \chi$$\parallel$$\bm Q$.
In this case, $\bm P$ should be zero within the spin-current model.
Thus we expect vanishing of $\bm P$ when we apply 
$\bm H_{\rm ex}$$\parallel$$\bm b$ ($Pbnm$ setting) to TbMnO$_3$ and 
DyMnO$_3$. 
However, reorientation of $\bm P$ from $\bm P$$\parallel$$\bm c$ to 
$\bm P$$\parallel$$\bm a$ is observed in reality 
[see Fig.~\ref{Fig01}(e)].
A neutron-scattering experiment confirmed that this $\bm P$ reorientation
results from the spin-chirality flop from $\bm \chi$$\parallel$$\bm a$ 
to $\bm \chi$$\parallel$$\bm c$~\cite{Aliouane09,Yamasaki08}.
This discrepancy has been naively attributed to the influence of $f$ 
moments on the rare-earth ions thus far~\cite{Prokhnenko07a,Prokhnenko07b}. 
However, a similar behavior has been observed also in 
LiCu$_2$O$_2$ without $f$ moments~\cite{SPark07}, 
suggesting an essentially new mechanism. Mostovoy reproduced the flop 
%%in $R$MnO$_3$ ($R$=Tb and Dy) 
by introducing higher-order anisotropies in a {\it phenomenological} theory 
%%based on the Landau free-energy expansion
although their microscopic origins are unclear~\cite{Mostovoy06}.
%%suggesting that an essentially new mechanism is necessary to explain 
%%this phenomenon.
On the other hand, the application of $\bm H_{\rm ex}$$\perp$$\bm Q$ is 
expected to stabilize a transverse conical spin order with 
$\bm \chi$$\perp$$\bm Q$. As shown in Fig.~\ref{Fig01}(d), 
we expect the $ab$-plane transverse conical order with 
$\bm P$$\parallel$$\bm a$
when we apply $\bm H_{\rm ex}$$\parallel$$\bm c$ to TbMnO$_3$ and DyMnO$_3$.
However, in TbMnO$_3$, the first-order transition 
to paraelectric ($\bm P$=0) phase is observed 
under $\bm H_{\rm ex}$$\parallel$$\bm c$ as shown in Fig.~\ref{Fig01}(f).
The $\bm H_{\rm ex}$-induced vanishing of $\bm P$ is also observed in 
MnWO$_4$~\cite{Taniguchi08} and Ni$_3$V$_2$O$_4$~\cite{Kenzelmann06}.

To solve these puzzles, we start with a classical Heisenberg model on a 
cubic lattice, in which the Mn $S$=2 spins are treated as classical vectors. 
The Hamiltonian 
%%contains frustrating spin exchanges, the single-ion spin
%%anisotropy, the DM interaction, and the Zeeman coupling, which is
is given by 
$\mathcal{H}=\mathcal{H}_J+\mathcal{H}_{\rm sia}
+\mathcal{H}_{\rm DM}+\mathcal{H}_{\rm Zeeman}$.
The first term 
$\mathcal{H}_J=\sum_{<i,j>} J_{ij} \bm S_i \cdot \bm S_j$ 
describes spin-exchange interactions as shown in Fig.~\ref{Fig01}(a).
The second term $\mathcal{H}_{\rm sia}$ denotes the single-ion spin
anisotropy, which consists of two parts as 
$\mathcal{H}_{\rm sia}=\mathcal{H}_{\rm sia}^D+\mathcal{H}_{\rm sia}^E$ 
with $\mathcal{H}_{\rm sia}^D=D\sum_{i} S_{\zeta i}^2$
and $\mathcal{H}_{\rm sia}^E=E\sum_{i} (-1)^{i_x+i_y}
(S_{\xi i}^2-S_{\eta i}^2)$.
%%%%%%%%%%%%%%%%%%%%%%%%%%%%%%%%%%%%%%%%%%%%%%%%%%%%%%%%%%%%%%%%%%
%%\begin{eqnarray}
%%\mathcal{H}_{\rm sia}^D&=&
%%D\sum_{i} S_{\zeta i}^2, \\
%
%%\mathcal{H}_{\rm sia}^E&=&E\sum_{i} (-1)^{i_x+i_y}
%%(S_{\xi i}^2-S_{\eta i}^2).
%%\end{eqnarray}
%%%%%%%%%%%%%%%%%%%%%%%%%%%%%%%%%%%%%%%%%%%%%%%%%%%%%%%%%%%%%%%%%%
Here $\xi_i$, $\eta_i$ and $\zeta_i$ are the tilted local axes attached 
to the $i$th MnO$_6$ octahedron~\cite{Alonso00}. 
The term $\mathcal{H}_{\rm sia}^D$ causes the hard-axis anisotropy along 
$\bm c$, or, equivalently, the easy-plane anisotropy in the $ab$ plane.
The third term 
$\mathcal{H}_{\rm DM}=\sum_{<i,j>}\bm d_{i,j}\cdot(\bm S_i \times \bm S_j)$ 
represents the DM interaction where
the vectors $\bm d_{i,j}$ are defined on the Mn($i$)-O-Mn($j$) bonds, 
and are expressed by five DM parameters, $\alpha_{ab}$, $\beta_{ab}$, 
$\gamma_{ab}$, $\alpha_c$, and $\beta_c$~\cite{Solovyev96}.
%%, because of the crystal symmetry~\cite{Solovyev96}.
This term consists of two parts, $\mathcal{H}_{\rm DM}^{ab}$ and
$\mathcal{H}_{\rm DM}^c$, where $\mathcal{H}_{\rm DM}^{ab}$ 
($\mathcal{H}_{\rm DM}^c$) is associated with the DM vectors on the in-plane
(out-of-plane) Mn-O-Mn bonds.
The last term, 
$\mathcal{H}_{\rm Zeeman}=g\mu_{\rm B} \sum_{i}\bm S_i \cdot \bm H_{\rm in}$, 
stands for the Zeeman coupling.
Note that the Mn spins feel the internal magnetic field $\bm H_{\rm in}$,
which consists of two contributions, i.e., the applied field $\bm H_{\rm ex}$
and the effective field $\bm H_{fd}$ from the $f$ moments.
This model has successfully reproduced the phase diagrams of $R$MnO$_3$
at $\bm H_{\rm ex}$=0~\cite{Mochizuki09b}.
%%in the absence of magnetic field~\cite{Mochizuki09b}.

%%%%%%%%%%%%%%%%%%%%%%%%%%%%%%%%%%%%%%%%%%%%%%%%%%%%%%%%%%%%%%%%%%
%%\begin{table}
%%\caption{Model parameters used in the calculations. The energy unit 
%%is meV.}
%%\begin{tabular}{c|c|cc}
%%\hline
%%\hline
%% & & TbMnO$_3$ & DyMnO$_3$  \\
%%\hline
%%$\mathcal{H}_{\rm ex}$  & ($J_{ab}$, $J_b$, $J_c$) 
%%& (-0.74, 0.64, 1.0) & (-0.70, 0.99, 1.0)  \\
%%$\mathcal{H}_{\rm sia}$ & ($D$, $E$) & (0.20, 0.25) &  (0.22, 0.25)  \\
%%$\mathcal{H}_{\rm DM}^{ab}$  
%%& ($\alpha_{ab}$, $\beta_{ab}$, $\gamma_{ab}$)
%%& (0.1, 0.1, 0.14) & (0.1, 0.1, 0.14)  \\
%%$\mathcal{H}_{\rm DM}^c$  & ($\alpha_c$, $\beta_c$)
%%& (0.48, 0.1) & (0.45, 0.1)  \\
%%\hline
%%\hline
%%\end{tabular}
%%\label{tabl:PRMVL}
%%\end{table}
%%%%%%%%%%%%%%%%%%%%%%%%%%%%%%%%%%%%%%%%%%%%%%%%%%%%%%%%%%%%%%%%%%
We have microscopically determined the values of $J_{ab}$, $J_b$, $J_c$, 
$D$, and $E$, and have estimated the values of five DM parameters 
in Ref.~\cite{Mochizuki09b}.
%%These parameters 
%%They are tuned within the range of uncertainty so as to 
%%reproduce the experimentally observed spin structures.
We perform calculations using two sets of the model parameters (A and B)
as 
(A) ($J_{ab}$, $J_b$, $J_c$)=($-$0.74, 0.64, 1.0), ($D$, $E$)=(0.2, 0.25),
($\alpha_{ab}$, $\beta_{ab}$, $\gamma_{ab}$)=(0.1, 0.1, 0.14) and
($\alpha_c$, $\beta_c$)=(0.48, 0.1), 
and 
(B) ($J_{ab}$, $J_b$, $J_c$)=($-$0.7, 0.99, 1.0), ($D$, $E$)=(0.22, 0.25),
($\alpha_{ab}$, $\beta_{ab}$, $\gamma_{ab}$)=(0.1, 0.1, 0.14) and
($\alpha_c$, $\beta_c$)=(0.45, 0.1).
Here the energy unit is meV.
These parameter sets give the $bc$-plane spin spirals
propagating along the $b$ axis with wave numbers $Q_b$=0.3$\pi$ and 
$Q_b$=0.4$\pi$, respectively. 
They reproduce well the spiral spin states 
in TbMnO$_3$ ($Q_b$=0.28$\pi$)~\cite{Kenzelmann05} 
and DyMnO$_3$ ($Q_b$=0.39$\pi$)~\cite{Kimura05} at $H_{\rm ex}$=0.

%%We study thermodynamic properties of this model 
We analyze this model 
using the replica-exchange MC technique~\cite{Hukushima96}. 
Each exchange sampling is taken after 400 standard MC steps. 
After 600 exchanges for thermalization, we typically perform 1000 
exchanges for systems of $N$=40$\times$40$\times$6 sites 
%%along the $x$, $y$ and $z$ axes 
with periodic boundaries.

%%%%%%%%%%%%%%%%%%%%%%%%%%%%%%%%%%%%%%%%%%%%%%%%%%%%%%%%%%%%%
\begin{figure}[tdp]
\includegraphics[scale=1.0]{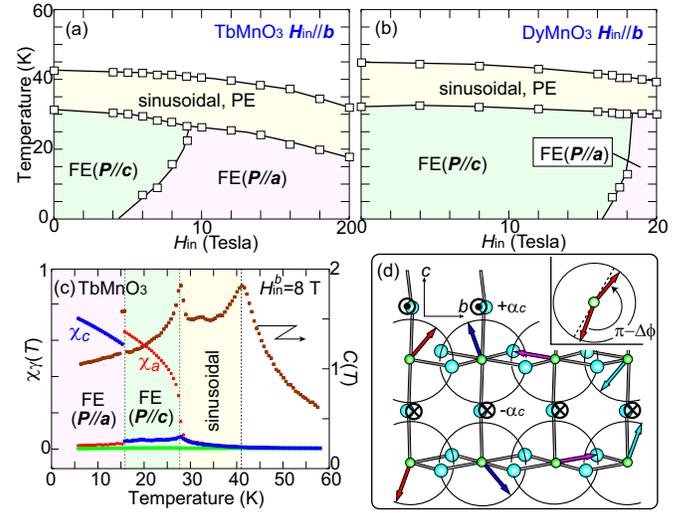}
\caption{(color online). Theoretical $T$-$H_{\rm in}$ phase diagrams of 
(a) TbMnO$_3$ and (b) DyMnO$_3$ for $\bm H_{\rm in}$$\parallel$$\bm b$. 
(c) $T$ profiles of specific heat $C(T)$ and spin chiralities 
$\chi_\gamma(T)$ ($\gamma$=$a$, $b$, $c$) for TbMnO$_3$ at 
$H_{\rm in}^b$=8 T.
(d) Spin structure in the $bc$-plane spiral state at $H_{\rm in}$=0, and 
arrangement of the $a$-axis components of DM vectors on the out-of-plane 
Mn-O-Mn bonds. The symbols $\odot$ and $\otimes$ express their signs, 
i.e., positive and negative, respectively. 
In the inset, the arrows (dashed lines) show the spin directions in the 
presence (absence) of DM interaction.}
\label{Fig02}
\end{figure}
%%%%%%%%%%%%%%%%%%%%%%%%%%%%%%%%%%%%%%%%%%%%%%%%%%%%%%%%%%%%%%
In Figs.~\ref{Fig02}(a) and (b) we display theoretically 
obtained $T$-$H_{\rm in}$ phase diagrams of TbMnO$_3$ and DyMnO$_3$ for 
$\bm H_{\rm in}$$\parallel$$\bm b$, respectively. 
They successfully reproduce the observed reorientation of 
$\bm P$ from $\bm P$$\parallel$$\bm c$ to $\bm P$$\parallel$$\bm a$ 
as a flop of the spin chirality from $\bm \chi$$\parallel$$\bm a$ 
to $\bm \chi$$\parallel$$\bm c$.
We determine the transition points and the spin structures by calculating
the $T$ dependence of specific heat 
$C(T)=\frac{1}{N} \partial\langle \mathcal{H}\rangle/\partial(k_{\rm B}T)$ 
and spin chiralities 
$\chi_\gamma(T)=\frac{1}{N} \langle |\sum_{i} 
(\bm S_i \times \bm S_{i+\hat b})_\gamma |\rangle/S^2$
($\gamma$=$a$, $b$, $c$).
%%%%%%%%%%%%%%%%%%%%%%%%%%%%%%%%%%%%%%%%%%%%%%%%%%%%%%%%%%%%%%
%%\begin{eqnarray}
%%C(T)&=&\frac{1}{N} \partial\langle \mathcal{H}\rangle/
%%\partial(k_{\rm B}T),\\
%% 
%%\chi_\gamma(T)&=&\frac{1}{N} \langle |\sum_{i} 
%%(\bm S_i \times \bm S_{i+\hat b})_\gamma |\rangle/S^2,
%%\end{eqnarray}
%%%%%%%%%%%%%%%%%%%%%%%%%%%%%%%%%%%%%%%%%%%%%%%%%%%%%%%%%%%%%%
Here the brackets denote thermal averages. 
Concerning the spin chiralities, the $\chi_a(T)$ [$\chi_c(T)$] 
has a large value, while other two components are nearly zero
in the $bc$-plane [$ab$-plane] spiral or conical phases. 
Figure~\ref{Fig02}(c) shows $C(T)$ and $\chi_\gamma(T)$ at 
$H_{\rm in}^b$=8 T for TbMnO$_3$. 
The $C(T)$ shows three peaks in accord with successive three 
phase transitions with lowering $T$. 
%%First, the system enters into the sinusoidal phase from the paramagnetic 
%%one, through which all of the three components $h_\gamma(T)$ are 
%%approximately zero constantly. 
%%At the second transition into the $bc$-plane spiral phase, 
%%the $a$-axis component $h_a(T)$ increases, while other two components 
%%remain to be nearly zero. 
%%With further lowering $T$, the $a$-axis component $h_a(T)$ 
%%suddenly drops, while the $c$-axis component $h_c(T)$ increases with 
%%a jump, indicating a spiral-plane flop from $bc$ to $ab$. 
The assignments of spin structures are confirmed by calculating 
spin and spin-chirality correlations in the momentum space.
%%calculated by the Fourier transformation of the spin and 
%%spin-helicity configuration.

%%%%%%%%%%%%%%%%%%%%%%%%%%%%%%%%%%%%%%%%%%%%%%%%%%%%%%%%%%%%%
\begin{figure}[tdp]
\includegraphics[scale=1.0]{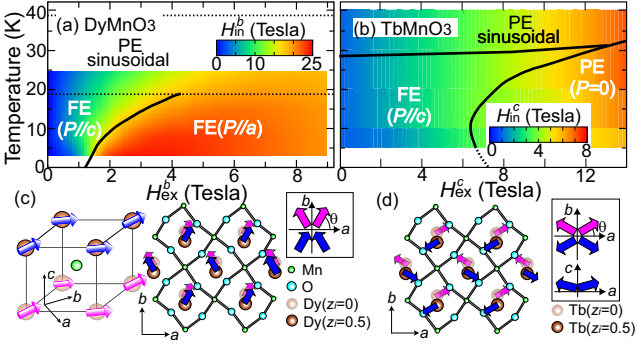}
\caption{(color). (a)[(b)] Intensity map of internal 
magnetic field $H_{\rm in}^b$ [$H_{\rm in}^c$] for DyMnO$_3$ [TbMnO$_3$]
in plane of $T$ and external magnetic field $H_{\rm ex}^b$ 
[$H_{\rm ex}^c$] calculated from experimental magnetization data 
$m_b(T,H_{\rm ex}^b)$ [$m_c(T,H_{\rm ex}^c)$], 
which reproduces the experimental $T$-$H_{\rm ex}$ diagram in 
Fig.~\ref{Fig01}(e)[(f)]. 
(c)[(d)] Arrangement of the Dy [Tb] $f$ moments in DyMnO$_3$ 
[TbMnO$_3$] under $\bm H_{\rm ex}$$\parallel$$\bm b$ 
[$\bm H_{\rm ex}$$\parallel$$\bm c$] where 
$\theta$$\sim$60$^{\circ}$ [$\theta$$\sim$30$^{\circ}$]~\cite{Kimura05}.}
\label{Fig03}
\end{figure}
%%%%%%%%%%%%%%%%%%%%%%%%%%%%%%%%%%%%%%%%%%%%%%%%%%%%%%%%%%%%%%
%%%%%%%%%%%%%%%%%%%%%%%%%%%%%%%%%%%%%%%%%%%%%%%%%%%%%%%%%%%%%
\begin{figure}[tdp]
\includegraphics[scale=1.0]{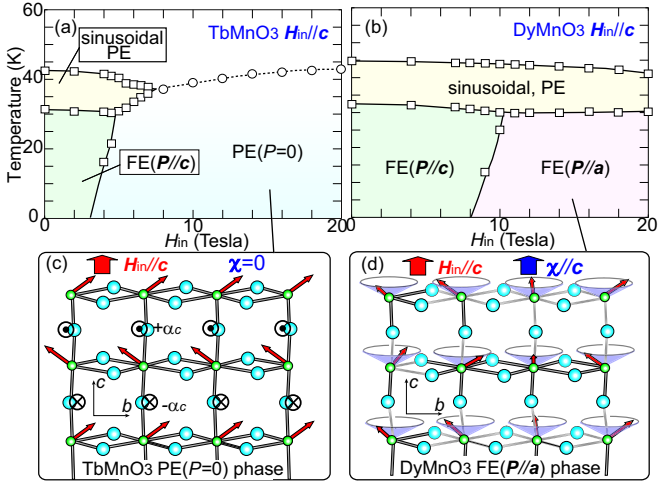}
\caption{(color online). Theoretical $T$-$H_{\rm in}$ phase diagrams of 
(a) TbMnO$_3$ and (b) DyMnO$_3$ for $\bm H_{\rm in}$$\parallel$$\bm c$, 
and spin structures in (c) PE($P$=0) phase in TbMnO$_3$ and 
(d) FE($\bm P$$\parallel$$\bm a$) phase in DyMnO$_3$.
Dashed line (open circles) in (a) denote the crossover line (points) 
where $C(T)$ shows a broad maximum.}
\label{Fig04}
\end{figure}
%%%%%%%%%%%%%%%%%%%%%%%%%%%%%%%%%%%%%%%%%%%%%%%%%%%%%%%%%%%%%%
By calculating the $H_{\rm in}$ dependence of the expectation value for 
each term in the Hamiltonian, we identify a mechanism of 
%%the transitions.
the chirality flop from $\bm \chi$$\parallel$$\bm a$ 
to $\bm \chi$$\parallel$$\bm c$ under 
$\bm H_{\rm ex}$$\parallel$$\bm b$.
%%can be understood as follows. 
The $bc$-plane spiral with $\bm \chi$$\parallel$$\bm a$ 
at $\bm H_{\rm ex}$=0 is stabilized by the DM interaction associated 
with the DM vectors on the out-of-plane Mn-O-Mn bonds, 
i.e., $\mathcal{H}_{\rm DM}^c$. 
The spins dominantly couple to the $a$-axis components of the vectors 
(i.e., components perpendicular to the $bc$ spiral plane)
whose signs are the same within a plane but alternate along the $c$ axis, 
while their magnitudes are all equal to $\alpha_c$ [see Fig.~\ref{Fig02}(d)]. 
Without DM interaction, angles between adjacent two spins along the $c$ axis 
are uniformly $\phi_c$=$\pi$ because of the strong antiferromagnetic (AFM) 
coupling $J_c$. In the presence of DM interaction, the angles 
%%become subject to a staggered modulation, i.e. 
alternate between $\pi+\Delta \phi_c$ and $\pi-\Delta \phi_c$ with 
$\Delta \phi_c$$>$0 [see the inset of Fig.~\ref{Fig02}(d)]. 
We can derive a gain of the DM energy due to this angle modulation as
$\Delta E_{\rm DM}^{bc}/N
=-\alpha_c S^2 |\cos\phi_c| \Delta\phi_c$.
%%%%%%%%%%%%%%%%%%%%%%%%%%%%%%%%%%%%%%%%%%%%%%%%%%
%%\begin{eqnarray*}
%%\Delta E_{\rm DM}^{bc}/N
%%&=&-\alpha_c S^2|\sin(\phi_c-\Delta\phi_c)-\sin\phi_c| \\
%%&=&-\alpha_c S^2 |\cos\phi_c| \Delta\phi_c.
%%\end{eqnarray*}
%%%%%%%%%%%%%%%%%%%%%%%%%%%%%%%%%%%%%%%%%%%%%%%%%%
Without $\bm H_{\rm in}$, the gain $\Delta E_{\rm DM}^{bc}$ in the 
$bc$-plane spiral dominates over the easy-($ab$)-plane 
[or the hard-($c$)-axis] spin anisotropy from $\mathcal{H}_{\rm sia}^D$, 
which favors the $ab$-plane spiral with $\bm \chi$$\parallel$$\bm c$. 
Note that the value of $|\cos\phi_c|$ is maximum (=1) at 
$\phi_c$=~$\pi$, but decreases as $\phi_c$ decreases. 
This means that the application of $\bm H_{\rm ex}$$\parallel$$\bm b$
suppresses this energy gain since it destroys the interplane 
AFM coupling and reduces the angle $\phi_c$ from $\pi$. 
The $bc$-plane spiral becomes destabilized when the reduced energy
gain $\Delta E_{\rm DM}^{bc}$ is defeated by the easy-($ab$)-plane
anisotropy $\mathcal{H}_{\rm sia}^D$, resulting in the 
spiral-plane (chirality) flop from $bc$ ($\bm \chi$$\parallel$$\bm a$)
to $ab$ ($\bm \chi$$\parallel$$\bm c$).
%%In the calculated $H$-dependence of the expectation values 
%%$\langle \mathcal{H}_{\rm sia}^D \rangle$ and 
%%$\langle \mathcal{H}_{\rm DM}^c \rangle$ (not shown), we indeed find
%%that the energy $\langle \mathcal{H}_{\rm DM}^c \rangle$ abruptly 
%%increases with a jump at the spiral-plane flop transition from 
%%$bc$ to $ab$, while the energy $\langle \mathcal{H}_{\rm sia}^D \rangle$ 
%%exhibits a sudden decrease.
Note that in $R$MnO$_3$, the $ac$-plane spiral or conical is unfavorable. 
%%Recent first-principles calculations for TbMnO$_3$ also confirmed this 
%%tendency~\cite{HJXiang08,Malashevich08}. 
This is because it can energetically benefit neither from 
$\mathcal{H}_{\rm sia}^D$ nor from $\mathcal{H}_{\rm DM}^c$, whereas the 
$ab$- and $bc$-plane spirals can take advantage of one of 
these two. We expect that the above mechanism is relevant also to the 
$\bm H_{\rm ex}$$\parallel$$\bm b$ induced $\bm P$ flop from 
$\bm P$$\parallel$$\bm c$ to $\bm P$$\parallel$$\bm a$
in LiCu$_2$O$_2$~\cite{SPark07} in terms of the role of 
$\bm H_{\rm ex}$, which destabilizes the spin spiral with 
$\bm P$$\parallel$$\bm c$~\cite{Kobayashi09} through 
destroying the AFM coupling along $\bm c$. 
Note that the single-ion anisotropy $\mathcal{H}_{\rm sia}^D$ cannot 
work in this quantum $S$=$1/2$ spin system in contrast to $R$MnO$_3$ 
with $S$=2 spins. We expect that the spin spiral with 
$\bm P$$\parallel$$\bm a$ under $\bm H_{\rm ex}$ 
(possibly the $ab$-plane spiral) is stabilized by the other 
interaction, and the DM coupling with the $c$-axis components 
of DM vectors is a possible candidate.

Now we compare our results with experimental ones. 
Between Figs.~\ref{Fig01}(e) and Fig.~\ref{Fig02}(b), 
there are a few discrepancies. First, threshold fields
for the $\bm P$ reorientation are different; i.e., the calculated threshold 
value of $H_{\rm in}^b$ for DyMnO$_3$ is approximately 18 T, 
whereas the experimental value of $H_{\rm ex}^b$ is 1-4 T. 
Second, the slope of the phase boundary is very steep in the theoretical 
$T$-$H_{\rm in}$ diagram of Fig.~\ref{Fig02}(b), while in the experimental 
$T$-$H_{\rm ex}$ diagram of Fig.~\ref{Fig01}(e), it is rather gradual. 
These discrepancies are solved by considering the effective magnetic 
field $\bm H_{fd}$ generated by the rare-earth $f$ moments, which acts 
on the Mn spins via the $f$-$d$ coupling $J_{fd}$.
Because of the AFM $J_{fd}$, $\bm H_{fd}$ and $\bm H_{\rm ex}$ are 
antiparallel, and the internal field $H_{\rm in}^\gamma$ 
($\gamma$=$a$, $b$, $c$) is given by 
$H_{\rm in}^\gamma=H_{\rm ex}^\gamma-H_{fd}^\gamma$.
Here $H_{fd}^\gamma$ is written using the $f$-electron 
magnetization $m_\gamma$ as a function of $T$ and 
$H_{\rm ex}^\gamma$ as 
$H_{fd}^\gamma(T,H_{\rm ex}^\gamma)=zJ_{fd}m_\gamma(T,H_{\rm ex}^\gamma)$.
%%%%%%%%%%%%%%%%%%%%%%%%%%%%%%%%%%%%%%%%%%%%%%%%%%
%%\begin{equation}
%%H_{fd}^b(T,H_{\rm ex}^b)=zJ_{fd}m_b(T,H_{\rm ex}^b).
%%\end{equation}
%%%%%%%%%%%%%%%%%%%%%%%%%%%%%%%%%%%%%%%%%%%%%%%%%%
Here $z$(=8) is the coordination number of $R$ ions around the Mn ion. 
We assume $J_{fd}$=0.45 T/$\mu_{\rm B}$ for DyMnO$_3$.
Figure~\ref{Fig03}(a) displays a color plot of the internal magnetic 
field $H_{\rm in}^b$ in the $T$-$H_{\rm ex}^b$ plane calculated using 
the experimental magnetization data. 
A solid line on which $H_{\rm in}^b$ is equal to the calculated 
threshold value is drawn. 
This figure coincides with the experimental diagram of DyMnO$_3$ 
in Fig.~\ref{Fig01}(e).
A similar analysis for TbMnO$_3$ has also reproduced the experimental 
diagram (not shown). The roles of the $f$-$d$ coupling in $R$MnO$_3$
at $H_{\rm ex}$=0 have been studied by recent neutron-scattering 
experiments~\cite{Prokhnenko07a,Prokhnenko07b}.
%%We find that the coupling also plays a role on the switching behaviors of 
%%$\bm P$ under $\bm H_{\rm ex}$.
We find that the switching of $\bm P$ can be qualitatively understood 
even without considering the $f$-$d$ coupling, but it should be taken into 
account for quantitative discussion.

Next we discuss the case of $\bm H_{\rm ex}$$\parallel$$\bm c$. 
The theoretical $T$-$H_{\rm in}^c$ phase diagrams of 
TbMnO$_3$ and DyMnO$_3$ are displayed in Figs.~\ref{Fig04}(a) and (b).
In Fig.~\ref{Fig04}(a), we find the transition to a coplanar spin state 
with $\bm P$=0 for TbMnO$_3$ at $H_{\rm in}^c$$\sim$3-5 T, 
which coincides with the experimental observation of paraelectric phase 
under $\bm H_{\rm ex}$$\parallel$$\bm c$. 
For its magnetic structure, see Fig.~\ref{Fig04}(c).
Again, there are a few discrepancies between the theoretical and 
experimental results [compare Figs.~\ref{Fig01}(f) and Fig.~\ref{Fig04}(a)].
%%concerning the phase boundary and range of the sinusoidal spin phase.
They are resolved by considering the influence of Tb $f$ moments.
In Fig.~\ref{Fig03}(b), we display the $T$ and 
$H_{\rm ex}^c$ dependence of the internal field $H_{\rm in}^c$ calculated 
from the experimental magnetization data. 
Here we assume $J_{fd}$=0.65 T/$\mu_{\rm B}$ for 
TbMnO$_3$. Solid lines on which $H_{\rm in}^c$ is equal to the calculated 
threshold value are drawn. This figure coincides well with the experimental 
diagram of TbMnO$_3$ in Fig.~\ref{Fig01}(f).
On the other hand, the transition to the $ab$-plane transverse conical state 
with $\bm P$$\parallel$$\bm a$ [see Fig.~\ref{Fig04}(d)] is found for 
DyMnO$_3$ in Fig.~\ref{Fig04}(b), which has not been observed in experiments 
up to $H_{\rm ex}^c$=9 T. The required $H_{\rm ex}^c$ 
for this transition deviates from the calculated critical value of 
$H_{\rm in}^c$ by the field $\bm H_{fd}$ from the 
Dy $f$ moments antiparallel to $\bm H_{\rm ex}$.
Hopefully, the reorientation of $\bm P$ will be observed in DyMnO$_3$ under 
a higher $H_{\rm ex}^c$. 

The contrasting behaviors of $\bm P$ under 
$\bm H_{\rm ex}$$\parallel$$\bm c$ between DyMnO$_3$ and TbMnO$_3$ 
can be attributed to the difference in magnitude of the in-plane
spin-exchange $J_b$.
TbMnO$_3$ has much smaller $J_b$=0.64 meV than DyMnO$_3$ with 
$J_b$=0.99 meV. At $H_{\rm ex}$=0, the Mn spins form a spiral order 
to minimize the spin-exchange energy in both compounds. 
Once we apply $\bm H_{\rm ex}$$\parallel$$\bm c$, the ferromagnetic 
moment is induced along the $c$ axis, and hence rotating components of 
the spins become reduced.
Then in TbMnO$_3$ with a small $J_b$, the spiral 
and conical spin orders no longer take advantage of the spin exchanges 
under $\bm H_{\rm ex}$$\parallel$$\bm c$, resulting in the first-order 
transition to the coplanar state as shown in Fig.~\ref{Fig04}(c). 
This state can benefit from all of the large $a$-axis 
components of the DM vectors on the out-of-plane bonds, 
which are perpendicular to the coplanar spin plane.
The $\bm H_{\rm ex}$-induced ferroelectric-to-paraelectric transition with 
sudden vanishing of $\bm P$ has also been observed in many other 
spin-spiral multiferroics, e.g., Ni$_3$V$_2$O$_8$~\cite{Kenzelmann06} and
MnWO$_4$~\cite{Taniguchi08}.
%%impurity-doped CuFeO$_2$~\cite{Seki07}.
We expect that the above mechanism is relevant also to them.

In summary, we have theoretically studied the puzzling $T$-$H_{\rm ex}$ 
phase diagrams of the spin-spiral multiferroic $R$MnO$_3$ ($R$=Tb and Dy)
and have revealed new mechanisms for the magnetic control of $\bm P$ by 
analyzing a microscopic spin model using the MC technique. 
We have shown that the applied $\bm H_{\rm ex}$$\parallel$$\bm Q$
($\parallel$$\bm b$ in the present case) reduces the DM energy through 
modulating the interplane spin angles, and thereby controls a competition 
between $\mathcal{H}_{\rm DM}^c$ and other interaction 
($\mathcal{H}_{\rm sia}^D$ in the present case), which results in the 
spiral-plane or spin-chirality flop with reorientation of $\bm P$.
On the other hand, the applied $\bm H_{\rm ex}$$\perp$$\bm Q$
($\parallel$$\bm c$ in the present case) suppresses the 
spin-exchange energy through reducing the rotating components of 
spins, and thereby causes a competition between the spin exchanges 
$\mathcal{H}_{\rm ex}$
and other interaction ($\mathcal{H}_{\rm DM}^c$ in the present case).
As a result, the first-order transition from spiral 
to coplanar spin phases occurs in TbMnO$_3$ with a rather small $J_b$
accompanied by the sudden disappearance of $\bm P$. We have discussed that
the proposed mechanisms are also applicable to many other 
spin-spiral multiferroics. Additionally, we have found that
the experimental results can be quantitatively reproduced by 
considering the effective field $\bm H_{fd}$ from the rare-earth $f$ 
moments.

We thank Y. Tokura and N. Nagaosa for valuable discussions. 
MM thanks H. Murakawa and Y. Tokunaga for their experimental supports.
This work was supported by Grant-in-Aid (No.22740214) and 
G-COE Program (``Physical Sciences Frontier") from MEXT Japan, 
and Funding Program for World-Leading Innovative R$\&$D on Science and 
Technology (FIRST Program) from JSPS.

%%%%%%


\begin{thebibliography}{999}
%%030526 Discovery TbMnO3, Giant magnetocapacitance, 
%%H-T PhsDgm for TbMnO3(H//b)
%%Magnetic control of ferroelectric polarization
\bibitem{Kimura03a}T. Kimura $et$ $al$., Nature (London) {\bf 426}, 55 (2003).

%%Reviews
\bibitem{Reviews}
Y. Tokura, J. Magn. Magn. Mater. {\bf 310}, 1145 (2007);
S.-W. Cheong and M. Mostovoy, Nat. Mater. {\bf 6}, 13 (2007).
%%D. I. Khomskii, J. Magn. Magn. Mater. {\bf 306}, 1 (2006).

%%070418 Review for spiral magnets as magnetoelectrics
%%Spiral Magnets as Magnetoelectrics
\bibitem{Kimura07}T. Kimura, Annu. Rev. Mater. Res. {\bf 37}, 387 (2007).

%%041122 Phase Diagrams for R=Gd, Tb, Dy, (Heq0 and Hne0),
%%GMC, Almost Review
%%Magnetoelectric phase diagrams of orthorhombic RMnO3 (R=Gd, Tb, and Dy)
\bibitem{Kimura05}T. Kimura, G. Lawes, T. Goto, Y. Tokura, and A. P. Ramirez, Phys. Rev. B {\bf 71}, 224425 (2005).

\bibitem{Murakawa08}H. Murakawa $et$ $al.$, Phys. Rev. Lett. {\bf 101}, 197207 (2008).

%%060919 LiCu2O2 Discovery of FE
%%Ferroelectricity in an S=1/2 Chain Cuprate
\bibitem{SPark07}S. Park, Y. J. Choi, C. L. Zhang, and S.-W. Cheong, Phys. Rev. Lett. {\bf 98}, 057601 (2007).

%%071001 FE P Flop in MnWO4 Induced by a Magnetic Field
%%H-dep. of FE P and spin-lattice coupling in MF MnWO4
\bibitem{Taniguchi08}K. Taniguchi $et$ $al$., Phys. Rev. B {\bf 77}, 064408 (2008).

%%051013 Ni3V2O8 H-switching of P
%%Field dependence of magnetic ordering in Kagome-staircase compound Ni3V2O8
\bibitem{Kenzelmann06}M. Kenzelmann $et$ $al$., Phys. Rev. B {\bf 74}, 014429 (2006).

%%051026 Inverse DM mechanism
%%Ferroelectricity in Spiral Magnets
\bibitem{Mostovoy06}M. Mostovoy, Phys. Rev. Lett. {\bf 96}, 067601 (2006).

%%041209 Spin-current model (inverse DM mechanism)
%%Spin Current and Magnetoelectric Effect in Noncollinear Magnets
\bibitem{Katsura05}H. Katsura, N. Nagaosa, and A. V. Balatsky, Phys. Rev. Lett. {\bf 95}, 057205 (2005).

%%050802 Inverse DM mechanism
%%Role of the Dzyaloshinskii-Moriya interaction in multiferroic perovskites
%%\bibitem{Sergienko06a}I. A. Sergienko and E. Dagotto, Phys. Rev. B {\bf 73}, 0%%94434 (2006).

%%050412 Neutron diffraction, TbMnO3
%%bc-plane spiral spin order is relevant to FE in TbMnO3
%%Magnetic Inversion Symmetry Breaking and Ferroelectricity in TbMnO3
\bibitem{Kenzelmann05}M. Kenzelmann $et$ $al$., Phys. Rev. Lett. {\bf 95}, 087206 (2005).

%%090211 NSD, TbMnO3, H//b, P//a phase=ab cycloid
%%Magnetic Feld induced flop of cycloidal spin order in multiferroic 
%%TbMnO3: The magnetic structure of the P//a phase
\bibitem{Aliouane09}N. Aliouane $et$ $al$., Phys. Rev. Lett. {\bf 102}, 207205 (2009).

%%080507 Detailed spin structure in P//a phase in Gd0.7Tb0.3MnO3, Neutron
%%Cycloidal Spin Order in the a-Axis Polarized FE Phase of 
%%Orthorhombic Perovskite Manganite 
\bibitem{Yamasaki08}Y. Yamasaki $et$ $al$., Phys. Rev. Lett. {\bf 101}, 097204 (2008).

\bibitem{Kobayashi09}Y. Kobayashi $et$ $al.$, J. Phys. Soc. Jpn. {\bf 78}, 084721 (2009). 

%%060903 NPD, DyMnO3, f-d coupling, H=0
%%Enhanced FeE Polarization by Induced Dy Spin Order in MF DyMnO3
\bibitem{Prokhnenko07a}O. Prokhnenko $et$ $al.$, Phys. Rev. Lett. {\bf 98}, 057206 (2007).

%%070816 NSD, TbMnO3, f-d coupling, H=0
%%Coupling of Frustrated Ising Spins to the Magnetic Cycloid in MF TbMnO3
\bibitem{Prokhnenko07b}O. Prokhnenko $et$ $al.$, Phys. Rev. Lett. {\bf 99}, 177206 (2007).

%%990802 Structure data, NPD, RMnO3(R=Pr,Nd,Dy,Tb,Ho,Er,Y)
%%Evolution of the JT Distortion of MnO6 Octahedra in RMnO3 Perovskites
%%(R=Pr,Nd,Dy,Tb,Ho,Er,Y): A Neutron Diffraction Study
\bibitem{Alonso00}For their direction vectors, we use the 
structural data of TbMnO$_3$ and DyMnO$_3$; 
see J. A. Alonso, M. J. Mart\'inez-Lope, M. T. Casais, and 
M. T. Fern\'andez-D\'iaz, Inorg. Chem. {\bf 39}, 917 (2000).

%%951127 LDA+U calc. for DM params.
%%Crucial Role of the Lattice Distortion in the Magnetism of LaMnO3
\bibitem{Solovyev96}I. Solovyev, N. Hamada, and K. Terakura, Phys. Rev. Lett. {\bf 76}, 4825 (1996).

%%090508 Full paper for the theoretical study on the PDs with zero H
%%Microscopic Model and Phase Diagrams of the Multiferroic RMnO3
\bibitem{Mochizuki09b}M. Mochizuki, and N. Furukawa, Phys. Rev. B {\bf 80}, 134416 (2009).

%%951204 Exchange Monte-Carlo
\bibitem{Hukushima96}K. Hukushima and K. Nemoto, J. Phys. Soc. Jpn. {\bf 65}, 1604 (1996).

%%070205 MF in CuFe1-xAlxO2 H-induced vanishing of P
%%Impurity-doping-induced FE in the frustrated antiferromagnet CuFeO2 
%%\bibitem{Seki07}S. Seki $et$ $al$., Phys. Rev. B {\bf 75}, 100403 (2007).
\end{thebibliography}
\end{document}